\documentclass[11pt]{article}

\usepackage{latexsym}
\usepackage{graphicx}
\begin{document}
\title{Reconsidering gauge-Higgs continuity}
\author{Michael Grady\\
Department of Physics\\ State University of New York at Fredonia\\
Fredonia NY 14063 USA}
\date{\today}
\maketitle
\thispagestyle{empty}
\begin{abstract}
The 3-d Z(2) lattice gauge-Higgs theory 
is cast in a partial axial gauge leaving a residual 
Z(2) symmetry, global in two directions and local in one. 
It is shown both analytically and numerically that this
symmetry breaks spontaneously in the Higgs phase and
is unbroken in the confinement phase. Therefore
they must be separated everywhere by 
a phase transition, in contradiction to a theorem by Fradkin and Shenker.
It relied on a fully fixed 
unitary gauge, which prohibits this phase transition 
explicitly. Thus the unfixed gauge theory is not, in this case, 
equivalent to the unitary-gauge version.

\end{abstract}

This letter presents evidence for a hidden symmetry breaking in 3-d Z(2) lattice gauge-Higgs theory, 
made visible by a partial axial gauge fixing. The symmetry which breaks is simply a subgroup of the original 
gauge symmetry, which is fixed enough to invalidate Elitzur's theorem \cite{elitzur} but leaves unbroken $N$
``layered" Z(2) symmetries on the $N^3$ lattice, one for each plane perpendicular to the third axis. 
In this gauge the average Higgs field on each layer and the average third-dimension-pointing link 
(link-magnetization) on 
each layer become order parameters for this residual exact $\rm{Z}(2)^N$ symmetry. 
It is found that in the Coulomb phase, the link magnetization
takes on a non-zero expectation value, but the Higgs field does not. In the Higgs phase they both have nonzero
expectation values, which can be shown analytically. In the confinement phase it can be shown analytically that 
both expectation values vanish and the symmetry is unbroken there. Therefore, due to the symmetry difference,
these phases must be separated by a phase transition.
This is a surprising result because a well-known theorem by Fradkin and Shenker (FS)\cite{fs} states that the Higgs and 
confinement phases are analytically connected. This phase-continuity
has played an important role in attempts to understand the confinement mechanism. The FS
proof relies crucially on using a completely-fixed unitary gauge, in which all of the Higgs fields are set to unity.
Normally, any fixed gauge is considered equivalent to the unfixed gauge theory, with the partition function
differing only by an infinite constant. However, if a portion of the gauge symmetry itself breaks spontaneously (e.g. 
global gauge symmetry or the layered symmetry considered above) then they will not be equivalent, because 
of the loss of ergodicity at a phase transition. The unfixed theory will differ from the unitary-gauge theory
by a {\em different} infinite constant in the symmetry-broken phase than in the unbroken phase. Therefore,
the FS proof, valid in the unitary gauge, is not necessarily valid  for the unfixed or 
partially-fixed theories. In fact the presence of the phase transition mentioned above shows it 
is invalid. The change in ergodicity at the phase transition produces a sudden change in entropy which can be seen as
the source of non-analyticity.

The plan of this letter is as follows. First, the previously-known phase structure of the 
theory is reviewed including the FS
analyticity region and the Monte-Carlo results of Jongeward, Stack, and Jayaprakash\cite{stack}. Then it is 
shown that in partial
axial gauge along the $\beta=0$ line, the theory is equivalent to the one-dimensional Ising model (here
$\beta$ is the gauge coupling parameter). This places a point of non-analyticity at ($\beta =0$, $\beta _H = \infty$),
well within the FS analyticity region. The full theory is self-dual, so if there is only one phase transition
it must occur along the self-dual line. Thus the FS analyticity arguments can be safely used away
from it.  They are used to show that the order parameters are identically zero in the strong
coupling region (confined phase) and non-zero in the dual region (Higgs phase). This is 
sufficient to prove the two regions are everywhere separated by a phase boundary.  Monte Carlo simulations are 
then used to
explore the low-$\beta$ region where no phase transition was seen in Ref.~\cite{stack}. Using the new order
parameters a clear transition is seen near the self-dual line, which is most likely weakly first-order.
Finally the situation in the 
Coulomb phase, where the symmetry is also broken is discussed. The fact that the gauge theory apparently 
``breaks itself" (i.e. breaks the layered partially-global gauge symmetry) has interesting implications for continuum
gauge theories.

The theory under consideration is a 3-d classical lattice theory with action
\[
S=\beta \sum _{\vec{x},k,l} (1-U_{\Box kl} (\vec{x}))+ \beta _H 
\sum _{\vec{x},k} (1-\phi (\vec{x})U_k (\vec{x})\phi(\vec{x}+\hat{k})) .
\]
The $U_k$ are gauge fields lying on the links of the lattice, $k=1$ to 3, and the $\phi$ are Higgs fields
living on sites. $U_{\Box kl}$ is the standard  plaquette variable. 
They all take values $\pm 1$. There is a Z(2) local gauge symmetry, because the
action is invariant if both the Higgs field and all links attached to a given site are 
multiplied by $\pm 1$. The phase diagram, as determined in Ref.~3, is shown in Fig.~1.
The theory on the right-hand boundary is just the 3-d Ising model and on the lower boundary the
3-d Ising lattice gauge theory, its dual \cite{wegner}. The second-order
transitions on the edges continue into the phase diagram and meet at the self-dual line, which is given by 
$\beta _H = -\frac{1}{2} \ln (\tanh \beta )$.
Either before or at this meeting the transitions become first-order, and the single joined transition
continues up the self-dual line where the first-order transition was seen 
to disappear, around $\beta=0.62$ in an apparent critical point.
No transition was seen beyond this.
If a phase transition simply ends in a critical point, it cannot be symmetry breaking. However, the attached
Ising
transition is symmetry-breaking. Indeed, it is argued below that the entire transition is symmetry breaking,
with broken and unbroken phases everywhere separated by a phase transition.
This
transition most likely continues up the self-dual line (unless it splits again, but that seems unlikely).

The results of this paper are mostly for the partial axial gauge.  As usual, we fix all 1-direction links to unity,
except for the $x_1 = 0$ plane. On $x_1 =0$ all 2-direction links are fixed to unity except for when $x_2 = 0$.
In the normal maximal-tree axial gauge, a line of third-direction links at $x_1 = x_2 =0$ are also fixed, however we 
will leave that line unfixed, giving a residual Z(2) symmetry on each $x_3$=const. plane. Because discrete 
symmetries can break spontaneously in two dimensions, this should be enough gauge fixing to evade Elitzur's theorem,
and allow one to view
whether or not the remaining partially-global symmetry breaks or not. 
Consider first, the theory on the $\beta=0$ line in this gauge. For this argument one can relax even further and
fix only the 1-direction links. Only the Higgs interaction remains. Along the 1-direction, due to the fixing
of the gauge field, one has a 1-d Ising interaction. The sideways 2 and 3-direction interactions do not 
affect the Higgs fields, because whatever the Higgs field configuration, one choice of sideways link will
always be frustrated and the other will always be satisfied. Because the sideways links are otherwise
unrestricted (due to the lack of a gauge interaction), when they are integrated out one has an equal contribution 
to every Higgs configuration. Therefore these interactions do not affect the Higgs field probability, and the
system is equivalent
to a set of 1-d Ising models. The 1-d Ising model is unbroken at non-zero temperature, but broken at $T=0$ 
($\beta _H = \infty$). The free energy has an essential singularity
here.  This is consistent with the requirement that if there is only one
line of singularities, it must be the self-dual line, which has ($\beta = 0 $, $\beta _H = \infty$)
as its endpoint. This singularity is within the FS analyticity zone. In the unitary gauge
in which the FS proof is given, all Higgs fields are set to unity, so the considered phase transition cannot take place
because the symmetry that is breaking is explicitly broken. This would seem to indicate that 
the proof is only valid in the unitary gauge.

Assuming the only possible singularities in the low-$\beta$ region are along the self-dual line, it seems safe
to assume the FS analyticity arguments are valid away from that line. Therefore, a convergent strong-coupling expansion
exists in the neighborhood of (0,0), and a convergent weak coupling expansion exists in the opposite dual region
($ \infty$, $\infty$). 
Consider calculating $< \phi >$ in the strong coupling expansion. Here the exponential in the partition
function is expanded in a power series, resulting in a terms involving the field $\phi _i$ and
various powers of terms from the action. The only terms which survive the partition-function summation
are those that have all even powers of fields. However, it is easily seen that for $<\phi _i>$ there will
always be an odd power of one or more $\phi$ fields. Therefore $< \phi > =0$ everywhere the strong coupling
expansion converges, i.e.\ at least in a finite patch around (0,0). Turning to the upper right hand corner of
the phase diagram, the $\beta= \infty$ line in the axial gauge is just the 3-d Ising model, which definitely has 
$<\phi > \neq 0$ in the high $\beta _H$ (Higgs) phase. 
A function identically zero in one portion of the plane cannot become nonzero
without passing through a singularity. Therefore, because an exact symmetry is being broken here, 
the Higgs and confinement
phases must be everywhere separated by a phase transition. 

In the symmetry-broken region, the infinite lattice loses ergodicity and the partition function is restricted to 
a single symmetry-broken sector. In the unbroken region the system is ergodic. In the unfixed or 
partially fixed gauges, the gauge-copy factor multiplying the partition function suddenly changes at the transition,
giving an extra term of $N \ln (2)$ to $\beta F$ in the broken phase (here $F$ is the free energy).  Although this
does not, in the thermodynamic limit,
change the specific free energy, $f =F/N^3$, it can change one of its derivatives. 
This is because the change
occurs over an increasingly smaller beta-range as the lattice size becomes infinite. This range is of order
the $\beta $-shift, 
$\Delta \beta \propto N^{-1/\nu }.$
If one computes higher derivatives of $f$ with respect to $\beta$ through finite differences, 
then one will eventually find one that diverges
as $N \rightarrow \infty$. In the unitary gauge, only one gauge copy is kept on both sides of the transition, so 
the non-analyticity is prevented from forming. This can be seen as due to explicitly breaking a symmetry that
would otherwise break spontaneously. Thus one can see that the unitary gauge is not valid in the case that part
of the gauge symmetry itself breaks spontaneously. In this case one must be careful not to fix those symmetries 
that will naturally break spontaneously, if one wishes a theory which is physically equivalent to the unfixed theory.
The safest fixed gauge is therefore one which only fixes to the point of invalidating Elitzur's theorem but no further.
Because a discrete symmetry cannot break spontaneously at non-zero temperature in less than two dimensions, it would seem
that our partially-fixed axial gauge that fixes independently 
on planes meets this criterion, but some additional minor thinning
of the gauge fixing might be possible.

Monte Carlo simulations 
were performed to illustrate the behavior of these new order parameters near the self-dual line in the region
previously thought to be continuous. A clear symmetry breaking is seen around 
the self-dual points $\beta= 0.5$, $\beta _H = 0.385968$, and $\beta=0.25$, $\beta _H=0.703415$, which are both 
well into the region seen as continuous in Ref.~\cite{stack}. 
The Higgs coupling was held fixed and $\beta$ varied. For the first of these, at gauge-couplings below and 
including  $\beta = 0.48$, the Higgs field order
parameter (average Higgs field on 2-d layers) shows a Gaussian distribution around zero, consistent with an
unbroken symmetry. Above $\beta=0.5$ it becomes non-Gaussian (as measured by Binder cumulant), first flattening
and  eventually
picking up a clear peak away from zero around $\beta = 0.58$ (Fig.~2). 
At the intermediate coupling of 0.55, a triple peak structure is seen, suggestive of a first-order phase 
transition (Fig.~3). One would need data from larger lattices to be certain of this identification of the order.
For the self-dual point at 
$\beta = 0.25$, similar behavior is seen (Gaussian below 0.23 and highly peaked away from origin above 0.33).
The layered link magnetization order parameter behaves similarly to the layered Higgs field 
at these couplings (Fig.~4),
but at a given $\beta$ is slightly less strongly broken. 
Simulations were done with generally 150,000 sweeps after 20,000-100,000 equilibration
sweeps. Hot and cold starts agreed reasonably well up to $\beta =0.585$, except that cold-start
distributions were still asymmetric. At $\beta = 0.6$
metastabilities apparently prevent agreement
even with 100,000 equilibration sweeps.  This strong metastability is also highly suggestive of a first-order transition.  
A simulated annealing procedure or cluster algorithm may be necessary to study the strongly-broken
region accurately. 
Monte Carlo simulations were also used to verify that the layered link order parameter undergoes a symmetry breaking
in the pure gauge theory ($\beta _H =0$) at the location of the known phase transition between Coulomb and confinement
phases. 

These transitions are unusual in that there are $N$ separate symmetries breaking instead of the usual one. Thus $N$
tunneling events must take place for the lattice to go from the completely broken to the completely unbroken state.
These can be watched by monitoring the order parameters for all of the layers. (For the pure gauge theory, the link order
parameters are not sensitive to a global Z(2) gauge transformation, so there are only $N-1$ broken symmetries,
however the link order parameter also breaks the Z(2) Polyakov loop symmetry in the third direction, giving 
a total of $N$). The total latent heat of the entire transition is also split into $N$ small jumps, making
it difficult to measure and see on top of random fluctuations. Thus without the order parameter
for guidance, the transition may be hard to distinguish from a crossover.  The effective volume that enters the
finite size scaling is for most quantities only $N^2$, the layer volume, 
as opposed to the total system size $N^3$. These features conspire to make this system a surprisingly difficult
one to study using the Monte Carlo method with finite-size scaling. Further details concerning
the finite-lattice scaling behavior will be given elsewhere.

In this letter it has been shown, both analytically and through Monte Carlo simulations, that the 3-d Z(2) 
gauge-Higgs system has a drastically different behavior in a partially-fixed axial gauge than it does in 
fully-fixed unitary gauge. This is because the partially-global Z(2) N-fold layered symmetry on an $N^3$
lattice breaks spontaneously in the Higgs (and Coulomb) phases, which is prevented from happening through 
explicit breaking in the unitary gauge. Therefore the Fradkin-Shenker proof of analyticity in a region encompassing
both the Higgs and confinement phases, performed in the unitary gauge, breaks down in the more relaxed
gauge, and by implication, in the unfixed case. The unitary gauge misses the singularity induced by the 
ergodicity change that occurs when the above symmetry-breaking takes place. In the partially-fixed axial
gauge it was shown that the Higgs and Confinement phases are everywhere separated by a phase transition.
The above arguments also hold in four dimensions and for continuous groups.
(here only in four dimensions due to the Mermin-Wagner theorem, since 3-d layers are needed to have a 
layered phase-transition at non-zero coupling for a continuous group).

The same order parameters (Higgs and link magnetizations) could be studied in the unfixed theory, by defining
a covariant Higgs field and covariant link in the following way. Simply multiply the Higgs field at a given
site by all of the links on the gauge-fixing tree from that site to the root for that layer. 
For the link, do this to
both ends.  These will be equal to the isolated Higgs field or link themselves
in the partial axial gauge. Then simply average these covariant objects over each layer to obtain order parameters
equivalent to our layered Higgs and link magnetizations. Assuming no further symmetry breakings take place, the
unfixed and partial axial gauge simulations will be equivalent. However, computing the nonlocal order 
parameters given above in the unfixed case would be numerically prohibitive. The idea that the known
phase transitions could be extended if nonlocal order parameters were considered was discussed in Ref.~\cite{sz}.

It is also very interesting to consider implications of the layered symmetry for the Coulomb phase, which was not
the focus of this paper. In the Coulomb phase, it is easy to show  that the 
Higgs order parameter is zero (due to the 3-d Ising transition on RHS of Fig.~1). However Monte Carlo simulations
show that the layered link magnetization order parameter is non-zero here, becoming zero at the confinement
transition. If this holds for continuous groups in 4-d, then for U(1) in four dimensions the layered 
3-d global 1-d local gauge symmetry is spontaneously broken at weak coupling. That is, 
the gauge symmetry is partially broken by the gauge field
itself.
This leads to a rather different picture in which gauge theories are much like magnetic spin
systems. The physical gauge fields (photons in the U(1) case) may be 
spin-wave like Goldstone collective excitations of the broken vacuum. Despite the vector order parameter,
Lorentz symmetry is effectively not broken because a combined Lorentz and gauge invariance remains unbroken.
Whether this realization simply results in a new point of view or has physical implications will 
require further study. Its also possible that certain gauge-fixings are not equivalent in the continuum
theory as well, because they might interfere with the formation of Goldstone modes. 
It is also very interesting to consider the new order parameters introduced here in the non-abelian case.
A preliminary investigation of the 4-d SU(2) lattice gauge theory shows the layered symmetry to be broken at 
weak coupling there as well\cite{mylayersu2}, which suggests
the pure SU(2) theory may itself have a Coulomb phase, contrary to usual expectations.

\newpage
\begin{figure}[ht]
                       \includegraphics[width=5in]{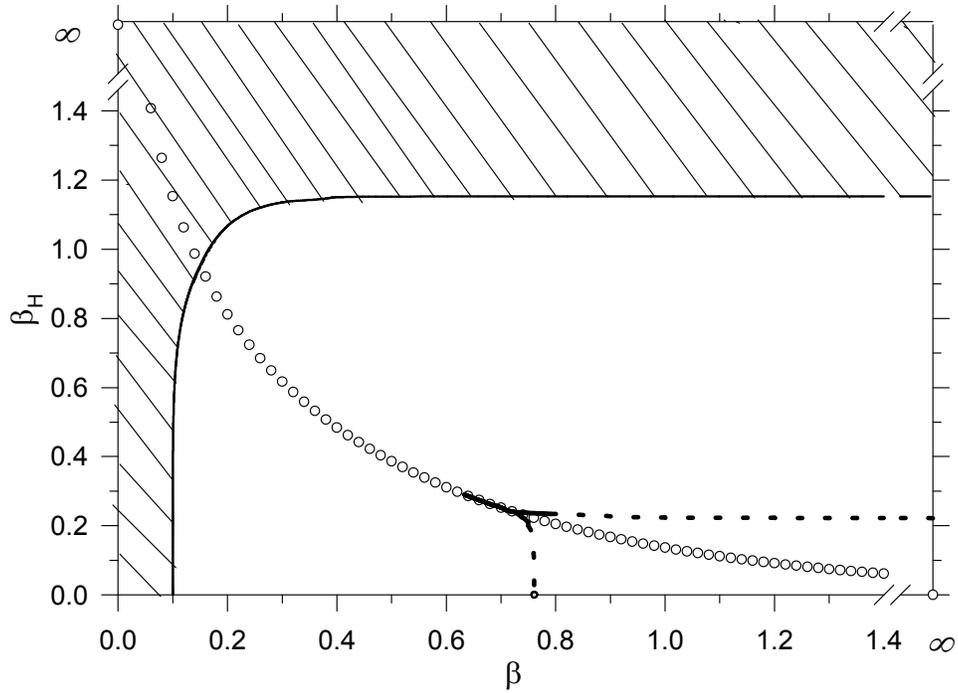}
            \caption{Previously known phase diagram of 3-d Z(2) gauge-Higgs model. Dashed lines are
second-order transitions, solid are first-order as found in Ref.~\cite{stack}. Open circles trace the self-dual line. 
Striped region is schematic representation of FS analyticity region (exact numerical bounds not given).}
          \label{fig1}
       \end{figure}
\newpage
\noindent
\begin{figure}[ht]
             \includegraphics[width=5in]{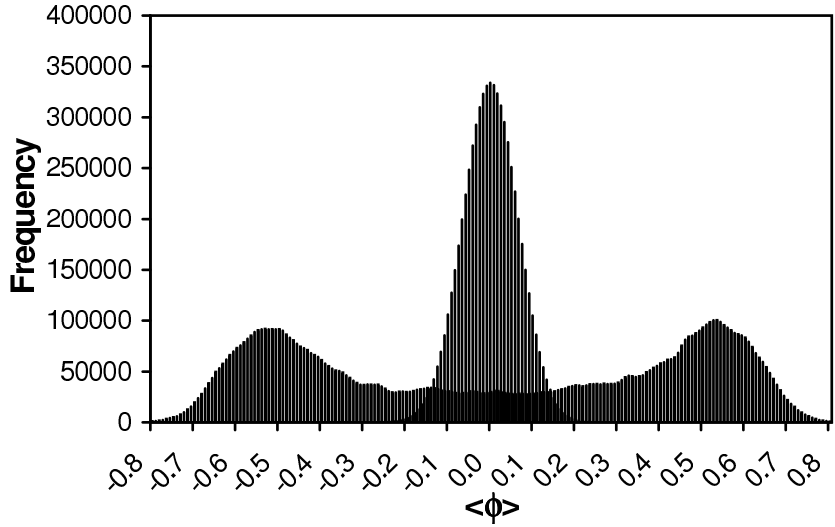} 
\caption{Histograms of Higgs field expectation value on lattice layers perpendicular to third direction on a $44^3$
lattice. Histogram centered on zero is for $\beta = 0.48$. Histogram with peaks away from
zero is for $\beta= 0.585$. $\beta _H$ is held fixed at $-\frac{1}{2}\ln (\tanh (0.5))$
for which $\beta =0.5$ would be the self-dual point. All runs shown here and later are hot starts. 
Bin size and spacing is carefully
chosen to have an equal number of possible
discrete values in each (8 in this case).}
\label{fig2}
\end{figure}
\begin{figure}[ht]
             \includegraphics[width=5in]{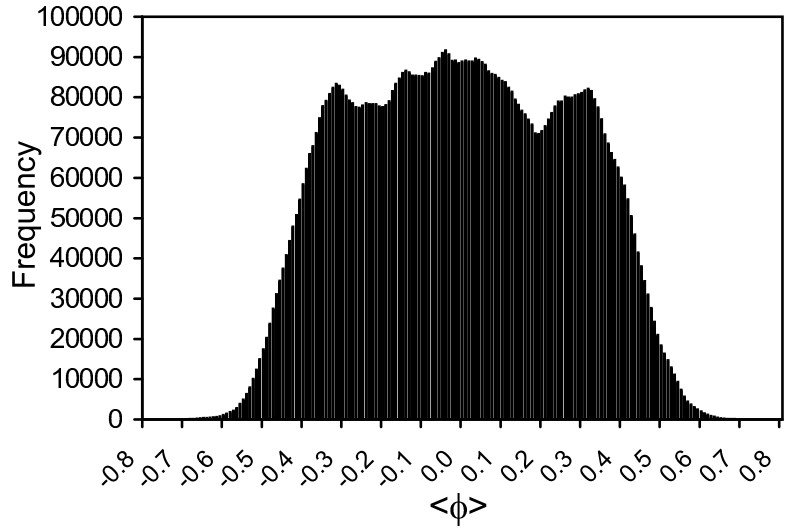}
\caption{Histogram of layered Higgs-field expectation value for $\beta = 0.55$ with same $\beta _H$.} 
\label{fig3}
\end{figure}
\begin{figure}[ht]
             \includegraphics[width=5in]{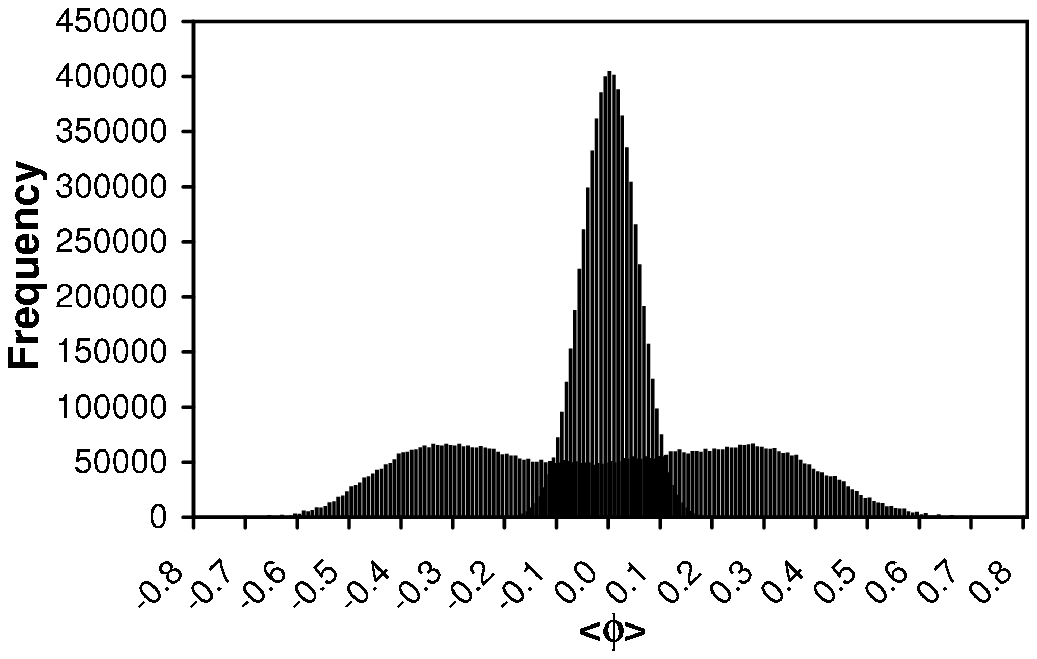}
\caption{Histogram of layered link expectation value for same $\beta$'s as Fig.~1.}
\label{fig4}
\end{figure}

\begin{thebibliography}{99}
\bibitem{elitzur}S. Elitzur, Phys. Rev. D {\bf 12}, 3978 (1975).
\bibitem{fs}E. Fradkin and S.H. Shenker, Phys. Rev. D {\bf 19}, 3682 (1979).
\bibitem{stack}G.A. Jongeward, J.D. Stack, and C. Jayaprakash, Phys. Rev. D {\bf 21}, 3360 (1980).
\bibitem{wegner}F.J. Wegner, J. Math. Phys. {\bf 12}, 2259 (1971).
\bibitem{sz}K. Szlachanyi, Commun. Math. Phys. {\bf 108}, 319 (1987).
\bibitem{mylayersu2}M. Grady, hep-lat/0409163. 
\end{thebibliography}
\end{document}